# Manipulation of polarization and spatial properties of light beams with chiral metafilms

V.V. Klimov,[1,2,3,7] I.V. Zabkov,[1,2,4] A.A. Pavlov,[1,2] R.-C.Shiu,[5] H.-C.Chan,[5] and G.Y.Guo[5,6,8]

[1]*Dukhov Research Institute of Automatics (VNIIA), Moscow 127055, Russia*
[2]*P.N. Lebedev Physical Institute, Russian Academy of Sciences, Moscow 119991, Russia*
[3]*National Research Nuclear University MEPhI, Moscow 115409, Russia*
[4]*Moscow Institute of Physics and Technology (State University), Dolgoprudny 141700, Russia*
[5]*Department of Physics, National Taiwan University, Taipei 10617, Taiwan*
[6]*Physics Division, National Center for ThereticalSciences,Hsinchu 300, Taiwan*
[7]*klimov256@gmail.com*
[8]*gyguo@phys.ntu.edu.tw*

**Abstract:** Two-dimensional lattices of chiral nanoholes in a plasmonic film with lattice constants being slightly larger than light wavelength are proposed for effective control of polarization and spatial properties of light beams. Effective polarization conversion and strong circular dichroism in non-zero diffraction orders in these chiral metafilms are demonstrated by electromagnetic simulations. These interesting effects are found to result from interplay between radiation pattern of single chiral nanohole and diffraction pattern of the planar lattice, and can be manipulated by varying wavelength and polarization of incoming light as well as period of metastructure and refractive indexes of substrate and overlayer. Therefore, this work offers a novel paradigm for developing planar chiral metafilm-based optical devices with controllable polarization state, spatial orientation and intensity of outgoing light.

## 1. Introduction

Control over spatial and polarization characteristics of light is important for many applications in optical devices. Nowadays miniaturization of such devices becomes crucial for such applications, and making use of plasmonic nanostructures seems to be promising in this direction. In particular, polarization conversion of electromagnetic waves by periodic nanostructures as well as by metasurfaces has been investigated theoretically and experimentally [1-5]. For example, Gordon *et. al.* observed strong polarization dependence in light transmission through a square lattice of nanoholes in metal on the ellipticity and orientation of the holes [1]. Fedotov *et. al.* demonstrated normal incidence transmission asymmetry of circularly polarized waves through a lossy anisotropic planar chiral structure [2]. Gansel *et. al.* reported significant extinction of circularly polarized plane waves by a square lattice of three-dimensional gold helices with the same handedness as of incident wave, while light waves of the other helicity pass the structure with high transmission [3]. In [4], asymmetrical transmission of linearly polarized light through a planar metamaterial composed of three-dimensional chiral meta-atoms with rotational symmetry was demonstrated. Finally, Zhao *et. al.* recently showed that planarized ultrathin broadband circular polarizers fabricated by planar technologies could show functionalities previously provided only by three-dimensional geometries [5].

Optical metamaterials considered in [1-5] have lattice constants being smaller than light wavelength. Consequently, only zero diffraction order is excited by incident light. In other words, these systems do not show diffraction effects. To control light more effectively, therefore, the concept of metasurfaces such as gradient metasurfases was recently put forward (see, e.g., [6-14]). The structure of gradient metasurfaces has two effective periods. One is the distance between nearest neighboring elements (meta-atoms) which is smaller than operating wavelength. The other effective period is the distance between similar groups of meta-atoms and is bigger than wavelength. Within this paradigm, metasurfaces which exhibit anomalous reflection and refraction which are nevertheless in agreement with generalized laws derived from Fermat's principle, was presented in [6]. Planar holey-metal lens made of concentric circular arrays of nanoscaled holes, which are used as a phase-shifting element, was demonstrated in [8]. Analog computing using reflective plasmonic metasurfaces was also reported recently in [9-11]. Metasurfaces that transform linearly polarized incident waves to circularly polarized outgoing waves in a wide wavelength range was demonstrated in [14].

We note that similar results can be obtained if subwavelength periodic features are replaced by individual meta-atoms (nanoparticles or nanoholes) of a special shape. As a result, interplay between radiation pattern of a single meta-atom and radiation pattern of the array can result in desirable behaviors of light beams. As an example of meta-atoms of sophisticated form, we can consider chiral meta-atoms with gammadion shape. This geometry is popular at present because it allows to effectively convert the polarization state of incident light (see, e.g., [15-20]).

In the present work, therefore, in order to control both polarization and spatial distribution of light beams, we consider, as an example, a square lattice of chiral nanoholes in a plasmonic film with the period being slightly larger than the wavelength (see Fig. 1) and analyze the polarization states of light scattered into different diffraction orders. To this end, we develop a formulation for evaluating polarization state-decomposed and diffraction channel-resolved transmissions and also perform numerical simulations for the metafilm with different sizes of chiral nanoholes. The rest of this article has the following structure. Section 2 contains basic formulae derived to analyze state of polarization and transmission coefficient of light wave scattered into various diffraction orders. In section 3, the results of numerical simulations for a planar lattice of gammadion-shaped nanoholes in a gold film with various nanohole sizes as well as calculated polarization conversion efficiency, circular dichroism and power spatial distribution are presented. Finally, conclusions drawn from this work are given in section 4.

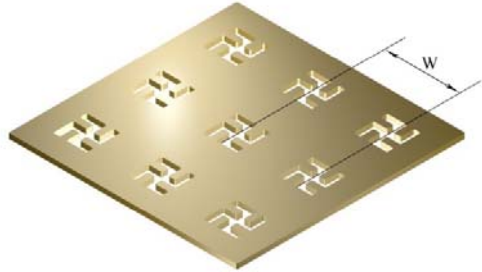

Fig. 1. A square lattice of chiral nanoholes in a gold layer. The period of the lattice is made slightly bigger that wavelength $W > \lambda$.

## 2. Analysis of polarization state of diffracted waves

Let us consider a square lattice of chiral nanoholes, which is excited by a normally incident plane wave (Fig. 2) with:

$$\mathbf{E}_{\text{inc}}(x, y, z) = \begin{pmatrix} E_{\text{inc},x} \\ E_{\text{inc},y} \\ 0 \end{pmatrix} \exp(ik^{\text{inc}} z), \quad (1)$$

where $k^{\text{inc}}$ is the absolute value of the incident wavevector in upper half space ($z > 0$) and

$E_{\text{inc},x}$ and $E_{\text{inc},y}$ are transverse components of the incident wave. Time dependence is assumed to be $\exp(i\omega t)$.

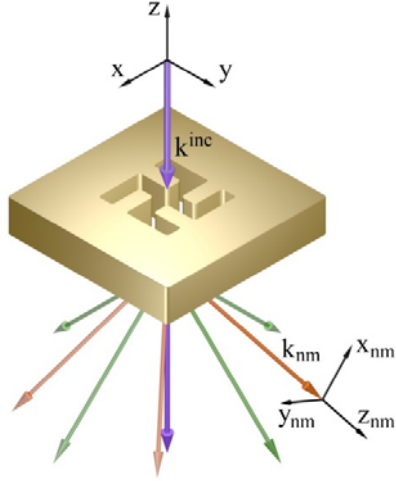

Fig. 2. Schematic of diffraction of a plane wave by a square lattice of chiral nanoholes in a gold film (Fig. 1) (only one unit cell is shown). The downward arrow above the film denotes the wavevector of incident field. The arrows below the film represent the wavevectors of diffracted waves of different orders. Global $x,y,z$ and local $x_{nm}, y_{nm}, z_{nm}$ coordinates system are shown.

Due to the periodicity of the system, the electric field of transmitted light can be expressed as sum of infinite number of plane waves:

$$\mathbf{E}(x,y,z) = \sum_{n=-\infty}^{\infty} \sum_{m=-\infty}^{\infty} \mathbf{C}_{nm} \exp(-ik_{x,n}x) \exp(-ik_{y,m}y) \exp(-ik_{z,nm}z) = \sum_{n=-\infty}^{\infty} \sum_{m=-\infty}^{\infty} \mathbf{E}_{nm}, \quad (2)$$

where $k_{x,n} = 2\pi n/W$, $k_{y,m} = 2\pi m/W$ and $k_{z,nm} = -\sqrt{k^2 - k_{x,n}^2 - k_{y,m}^2}$ are the components of the wavevector that correspond to $n,m$ diffraction order, $k$ is the absolute value of the wavevector in lower half space, $W$ is the lattice constant of the square lattice, and $x, y, z$ are the Cartesian coordinates. $\mathbf{E}_{nm}$ is the contribution to total electric field from the wave of $n,m$ diffraction order. The Fourier coefficients $\mathbf{C}_{nm}$ can be calculated from electric field in, e.g., the plane of $z = 0$ by:

$$\mathbf{C}_{nm} = \frac{1}{W^2} \int_0^W \int_0^W \mathbf{E}(x,y,0) \exp\left(i\left(k_{x,n}x + k_{y,m}y\right)\right) dx dy. \quad (3)$$

Let us write the wavevector of $n,m$ diffraction order as $\mathbf{k}_{nm} = \left(k_{x,n}, k_{y,m}, k_{z,nm}\right)^{\text{T}}$. To separate transmittances of the LCP and RCP waves, it is convenient to introduce a local coordinate system for each diffraction order. For this purpose, let us take the $z_{nm}$ axis of the local coordinate system for $n,m$ diffraction order to be parallel to wavevector $\mathbf{k}_{nm}$, i.e., $\mathbf{k}_{nm} \uparrow\uparrow \hat{\mathbf{z}}_{nm}$ (see Fig. 2). Transformation to such a coordinate system can be done by rotating by angle $\alpha = \arccos(k_{z,nm}/k)$ around vector $\mathbf{k}_{nm} \times \hat{\mathbf{z}}$, where $\hat{\mathbf{z}}$ is unit vector of the global

coordinate system. The corresponding rotation matrix is:

$$\mathbf{P}_{nm} = \begin{pmatrix} \dfrac{k_{x,n}^2 \cos\alpha + k_{y,m}^2}{k_{x,n}^2 + k_{y,m}^2} & \dfrac{k_{x,n}k_{y,m}(-1+\cos\alpha)}{k_{x,n}^2 + k_{y,m}^2} & -\dfrac{k_{x,n} \sin\alpha}{\sqrt{k_{x,n}^2 + k_{y,m}^2}} \\ \dfrac{k_{x,n}k_{y,m}(-1+\cos\alpha)}{k_{x,n}^2 + k_{y,m}^2} & \dfrac{k_{x,n}^2 + k_{y,m}^2 \cos\alpha}{k_{x,n}^2 + k_{y,m}^2} & -\dfrac{k_{y,m} \sin\alpha}{\sqrt{k_{x,n}^2 + k_{y,m}^2}} \\ \dfrac{k_{x,n} \sin\alpha}{\sqrt{k_{x,n}^2 + k_{y,m}^2}} & \dfrac{k_{y,m} \sin\alpha}{\sqrt{k_{x,n}^2 + k_{y,m}^2}} & \cos u \end{pmatrix}. \quad (4)$$

As a result, $\mathbf{k}'_{nm} = (0,\ 0,\ k'_{z,nm})^T = \mathbf{P}_{nm}\mathbf{k}_{nm}$ and $k'_{z,nm} = \sqrt{k_{x,n}^2 + k_{y,m}^2}\sin\alpha + k_{z,nm}\cos\alpha$. Here the prime represents the vector and its components in the local coordinate system. Clearly, in the local coordinate system, wavevector $\mathbf{k}'_{nm}$ has only the $z$ component. Electric field of the plane wave corresponds to the $n, m$ diffraction order is:

$$\mathbf{E}_{nm} = \mathbf{C}_{nm} \exp(-ik_{x,n}x)\exp(-ik_{y,m}y)\exp(-ik_{z,nm}z) = \begin{pmatrix} E_{x,nm} \\ E_{y,nm} \\ E_{z,nm} \end{pmatrix}, \quad (5)$$

In the local coordinate system, the electric field $\mathbf{E}'_{nm}$ can be written as:

$$\mathbf{E}'_{nm} = \mathbf{P}_{nm}\mathbf{E}_{nm} = \begin{pmatrix} (\mathbf{P}_{nm}\mathbf{C}_{nm})_x \\ (\mathbf{P}_{nm}\mathbf{C}_{nm})_y \\ 0 \end{pmatrix} \exp(-ik'_{z,nm}z_{nm}). \quad (6)$$

Equation (6) explicitly shows that the electric field in the local coordinate system is purely transverse. The corresponding Poynting vector is:

$$S_{z',nm} = \frac{1}{2Z}\left(|E'_{x,nm}|^2 + |E'_{y,nm}|^2\right) = \frac{1}{2Z}\left(|(\mathbf{P}_{nm}\mathbf{C}_{nm})_x|^2 + |(\mathbf{P}_{nm}\mathbf{C}_{nm})_y|^2\right), \quad (7)$$

where $Z = \sqrt{\mu_0/(\varepsilon_0 \varepsilon_{\text{bottom}})}$ is the wave impedance in the bottom half-space. $\mu_0$ and $\varepsilon_0$ are permeability and permittivity of vacuum, respectively. $\varepsilon_{\text{bottom}}$ is the relative permittivity of the bottom half-space and relative permeability in the same region is $\mu_{\text{bottom}} = 1$.

In order to separate transmittances to LCP and RCP waves, we introduce basis of right (R) and left (L) circularly polarized waves. Unit vectors of this basis, $\mathbf{e}_{R,nm}$ and $\mathbf{e}_{L,nm}$, are defined in terms of $\hat{\mathbf{x}}_{nm}$ and $\hat{\mathbf{y}}_{nm}$ of the local coordinate system as $\mathbf{e}_{R,nm} = (\hat{\mathbf{x}}_{nm} - i\hat{\mathbf{y}}_{nm})/\sqrt{2}$ and $\mathbf{e}_{L,nm} = (\hat{\mathbf{x}}_{nm} + i\hat{\mathbf{y}}_{nm})/\sqrt{2}$. This definition correspond to clockwise(counterclockwise) rotation of electric field vector from the point of view of source for RCP(LCP) wave. Therefore, the electrical field components of RCP and LCP waves for the $n, m$ diffraction order in the local coordinate system can be written as:

$$\mathbf{E}'_{nm} = E'_{R,nm}\mathbf{e}_{R,nm} + E'_{L,nm}\mathbf{e}_{L,nm}, \quad (8)$$

where $E'_{R,nm} = (E'_{x,nm} + iE'_{y,nm})/\sqrt{2}$ and $E'_{L,nm} = (E'_{x,nm} - iE'_{y,nm})/\sqrt{2}$. Using Eq. (8), we can find contribution to the Poynting vector for each diffraction order from waves with left and right circular polarizations. We should stress that rotation of the local coordinate system around $\hat{\mathbf{z}}_{nm}$ would affect only the values of $E'_{x,nm}$ and $E'_{y,nm}$, while $E'_{R,nm}$ and $E'_{L,nm}$ should remain unchanged.

By substituting Eq. (8) into Eq. (7), the Poynting vector can be written as:

$$S_{z',nm} = \frac{1}{2Z}\left(\left|E'_{R,nm}\right|^2 + \left|E'_{L,nm}\right|^2\right) = S_{R,nm} + S_{L,nm},$$
$$S_{R,nm} = \frac{1}{2Z}\left|E'_{R,nm}\right|^2, \quad S_{L,nm} = \frac{1}{2Z}\left|E'_{L,nm}\right|^2, \quad (9)$$

where $S_{R,nm}$ and $S_{L,nm}$ are the contributions to the total Poynting vector from RCP and LCP waves in the $n,m$ diffraction order. One can then define the corresponding transmission coefficients as the projections of $S_{R,nm}$ and $S_{L,nm}$ on the $z$ axis of the global coordinate system normalized to the Poynting vector of the incident wave $S_0$ as:

$$T^G_{P_I \to R,nm} = \left|k_{z,nm}/k\right| S_{R,nm}/S_0,$$
$$T^G_{P_I \to L,nm} = \left|k_{z,nm}/k\right| S_{L,nm}/S_0, \quad (10)$$

where $P_I = (R, L)$ denotes the circular polarization of incident light and $G = (R, L)$ denotes the handedness of gammadions in the square lattice (Fig. 1). For example, $T^R_{P_I \to (R,L),nm}$ corresponds to the transmission through a system with right twisted gammadions while $T^L_{P_I \to (R,L),nm}$ corresponds to the transmission through a system with left twisted gammadions.

### 3. Numerical results and discussion

Let us now apply the formalism described in the preceding section to the structure shown in Fig. 3. The gold film of 220 nm thickness is deposited on a quartz substrate and is then covered with immersion oil in the other side. The hole is also filled with immersion oil. The optical dielectric constants of gold from Weber [21] are used. Refractive indexes of quartz and oil are set to 1.443 and 1.51, respectively. Frequency dispersion of the dielectrics is neglected. The gammadion geometry is defined by a single parameter $s$, the size of the 17 identical squares which make up the gammadion, as shown in Fig. 3. Here we set $s = 86\,\text{nm}$. The edges of the squares sitting at the ends and corners of the gammadion are smoothed with curvature radius of $s/5$. The smoothing of edges is rather important because this makes the problem more definite from mathematical point of view and also increases the accuracy of calculations. For more detailed analysis of the problem of sharp edges, see [22].

The square lattice constant is taken to be $W = 1\,\mu m$. The wavelength considered is in the range of $\lambda = 760 - 820$ nm. The maximum order of diffraction peaks that appear in this system is $|n| = 1, |m| = 1$. The system is illuminated normally by a plane wave with either type of circular polarizations. Nanoholes with both left and right twists are considered. Twist of gammadions is defined from the point of view of incident wave. Right-handed (left-handed) gammadion has clockwise (counterclockwise) twist. In Fig.3, only the left twisted gammadion is shown.

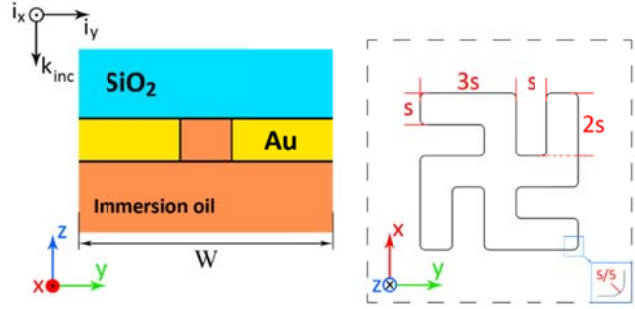

Fig. 3. Cross sections of the proposed square lattice of nanoholes in a gold film (Fig. 1). The gold film is mounted on the $SiO_2$ substrate. The other side of the film as well as the nanoholes are in immersion oil. The system is illuminated normally by plane waves from the $SiO_2$ side. The gammadion nanoholes have left twist from the viewpoint of incident light (or from the positive $z$ direction).

*3.1 Conversion of polarization state by nonzero diffraction orders*

The problem is solved numerically by the finite element method, as implemented in COMSOL Multiphysics. A square lattice of left twisted nanoholes irradiated with LCP and RCP waves is considered. Because of the $C_{4,z}$ symmetry of the system, diffraction orders with fixed $n^2 + m^2$ have the same transmittance $T^G_{P_I \to (R,L),nm}$, and hence one can rewrite the transmittances as:

$$\begin{aligned} U^0_{P_I \to (R,L)} &= T^L_{P_I \to (R,L),00}, \\ U^1_{P_I \to (R,L)} &= \sum_{n^2+m^2=1} T^L_{P_I \to (R,L),nm} = 4 T^L_{P_I \to (R,L),10}, \\ U^2_{P_I \to (R,L)} &= \sum_{n^2+m^2=2} T^L_{P_I \to (R,L),nm} = 4 T^L_{P_I \to (R,L),11}, \end{aligned} \quad (11)$$

to characterize the transmitted total power of symmetry-related orders. In Eq. (11), the L superscript of $T$ denotes the left twisted nanoholes. Figures 4 and 5 show, respectively, the wavelength dependence of transmittances $U^{n^2+m^2}_{L \to R}$ and $U^{n^2+m^2}_{L \to L}$ as well as $U^{n^2+m^2}_{R \to R}$ and $U^{n^2+m^2}_{R \to L}$ into different diffraction orders $n, m$. The same calculations have also been carried out for right twisted nanoholes (not shown here). Nevertheless, the results are found to be unaffected by simultaneously changing polarization state of the incident light and handedness of the nanoholes in the square lattice, as they should.

Figures 4 and 5 show that there is no polarization conversion in zero diffraction order in the proposed system (Figs. 1-3). For example, Fig. 4 clearly shows that when the system is illuminated by LCP light waves, zero diffraction order waves are purely LCP (i.e., there is no RCP component in the transmittance to the zero order). This conclusion is also confirmed by our further simulations for the same system using different values of the *s* parameter ($s = 43, 50, 65$ nm) (see Fig. 3). This is consistent with the statement reported in [23] that structures with the $C_{4,z}$ symmetry do not lead to the polarization conversion of circularly polarized light. In contrast, for other diffraction orders, significant polarization conversion occurs. We have also analyzed the polarization conversion property of nanoholes of different sizes (43, 50, 65 nm). Interestingly, we find that for all considered sizes of nanoholes, the conversion in orders $n^2 + m^2 = 2$ is more effective than in orders of $n^2 + m^2 = 1$ (see, e.g., Fig. 4). Furthermore, our simulations also reveal that bigger nanoholes give a more effective polarization conversion. Also, if the helicity of the incident wave matches the handedness of the nanoholes, the polarization conversion in orders $n^2 + m^2 = 2$ is more effective than the mismatched case. Indeed, Fig. 4 shows that when a LCP wave irradiates left twisted nanoholes, there is a wavelength range in which the diffracted waves of the $n^2 + m^2 = 2$ order, has more RCP content than that of LCP in the wavelength from ~783 to ~802 nm. And at $\lambda$ = 786 nm, the conversion of the LCP to RCP light is nearly perfect.

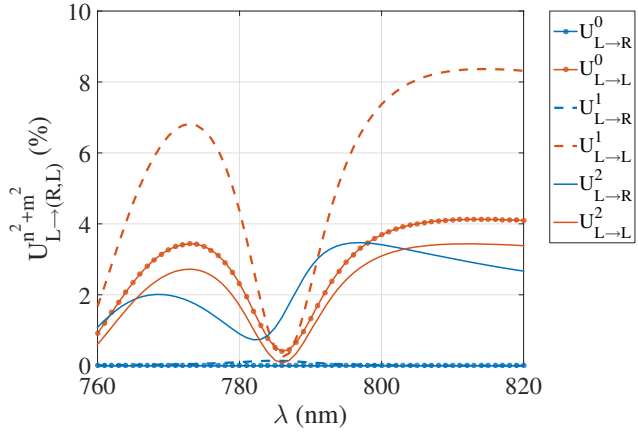

Fig. 4. Polarization state-decomposed transmittances $U_{L \to R}^{n^2+m^2}, U_{L \to L}^{n^2+m^2}$ (Eq. 11) versus vacuum wavelength for the considered diffraction orders (*n*, *m*) through a square lattice of left twisted nanoholes (Fig. 3) irradiated normally by a LCP plane wave. Blue and red lines represent RCP and LCP states of diffracted beams, respectively. Circles, dashed and solid lines correspond, respectively, to $n = m = 0$, $n^2 + m^2 = 1$ and $n^2 + m^2 = 2$.

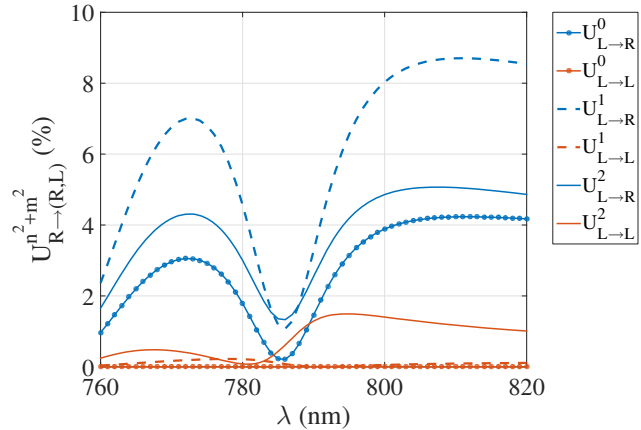

Fig. 5. Polarization state-decomposed transmittances $U_{L \to R}^{n^2+m^2}, U_{L \to L}^{n^2+m^2}$ (Eq. 11) versus vacuum wavelength for the considered diffraction orders (*n*, *m*) through a square lattice of left twisted nanoholes (Fig. 3) irradiated normally by a RCP plane wave. Blue and red lines represent RCP and LCP states of diffracted beams, respectively. Circles, dashed and solid lines lines correspond, respectively, to $n = m = 0$, $n^2 + m^2 = 1$ and $n^2 + m^2 = 2$.

To better understand this interesting effect, the asymmetry of LCP and RCP components versus wavelength in diffraction order $n^2 + m^2 = 2$ is displayed in Fig. 6. Here the polarization asymmetry is defined as

$$P_{nm}^{G} = \frac{T_{L \to R, nm}^{G} - T_{L \to L, nm}^{G}}{T_{L \to R, nm}^{G} + T_{L \to L, nm}^{G}} \qquad (12)$$

where the transmission coefficients $T_{L \to R, nm}^{G}, T_{L \to L, nm}^{G}$ are given by Eq. (10). Let us be reminded that superscript $G = L$ or $R$ denote left and right twists of the gammadions, respectively. It follows from Eq. (12) that $P_{nm}^{G} = 0$ corresponds to linearly polarized waves. $P_{nm}^{G} = +1$ and $P_{nm}^{G} = -1$ correspond to pure RCP and LCP waves, respectively. Two cases are considered here, namely, LCP wave irradiates either right or left twisted nanoholes.

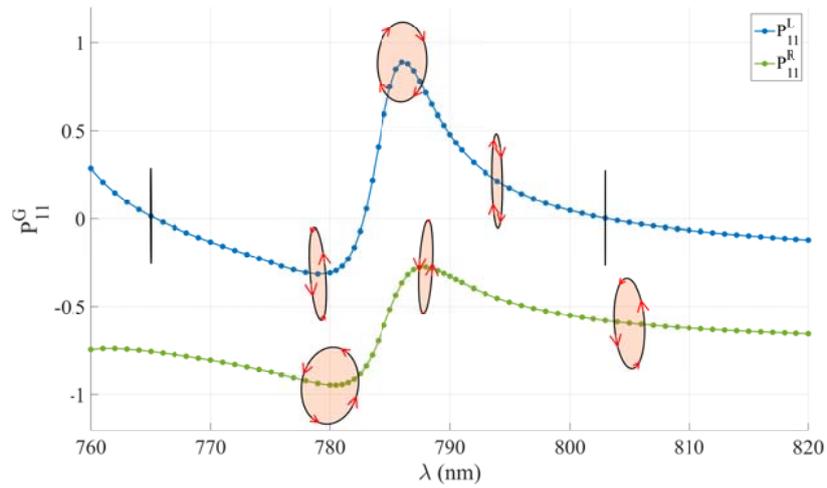

Fig. 6. Wavelength dependence of circular polarization asymmetry (Eq. 12) of light transmitted into $n^2 + m^2 = 2$ diffraction order through the square lattice (Fig. 3) made of left (blue line) and right (green line) twisted nanoholes irradiated by LCP plane wave. The ellipses denote the states of polarization at the wavelengths and the arrows on the ellipses indicate the helicities (or electric field vector) from the viewpoint of the incident light.

Figure 6 shows that when a LCP wave hits the square lattice of right twisted nanoholes, the polarization of waves transmitted to $n^2 + m^2 = 2$ diffraction orders changes with wavelength from nearly pure LCP to an elliptical one with left twist ($P_{11}^{R} < 0$), which matches the helicity of incident wave. On the other hand, when polarization of incident waves matches twist of gammadions (i.e., LCP wave and left twisted nanoholes), the polarization of waves transmitted to $n^2 + m^2 = 2$ diffraction orders changes from a highly elliptical one with left twist ($P_{11}^{L} < 0$) to an elliptical with right twist ($P_{11}^{L} > 0$). The polarization can even become almost pure RCP ($P_{11}^{L} = 0.9$) at $\lambda = 786$ nm, and this means nearly full conversion of incident LCP to RCP in $n = m = 1$ lobe. Moreover, the polarization can also become purely linear at wavelengths of 765 nm and 803 nm (Fig. 6). The results for all orders with $n^2 + m^2 = 2$ are the same.

The effect of polarization conversion in non-zero diffraction lobes is due to interplay between radiation patterns of single chiral nanohole and a square lattice of point sources. First of all, radiation pattern of single chiral nanohole has directions in which polarization of scattered light differs from polarization of incident light. This effect is of purely geometrical nature and, strictly speaking, can be observed even for light scattering by metallic sphere [24]. However, handedness of nanoholes leads to substantial increase of efficiency of polarization conversion. In particular, our simulations for the square lattice of cylindrical holes with the same cross-section area as that of $s = 86$ nm gammadions show that the effect is much weaker. For example, maximum value of $P_{nm}$ (Eq. 12) is equal to 0.13 at 799 nm wavelength while, for gammadion-shaped holes, it is 0.9 at 786 nm. Radiation pattern of the whole system can be found as multiplication of radiation pattern of single chiral nanohole with radiation pattern of a square lattice of point sources [25]. Therefore, when lobes of radiation pattern of the square lattice of point sources coincide with angle of maximum polarization conversion of single chiral nanohole, an effective polarization conversion into this lobe occurs. Interestingly, superposition of maximum of polarization conversion with lobes of radiation pattern can be tuned by changing wavelength $\lambda$, lattice constant $W$ and refractive index of surrounding media.

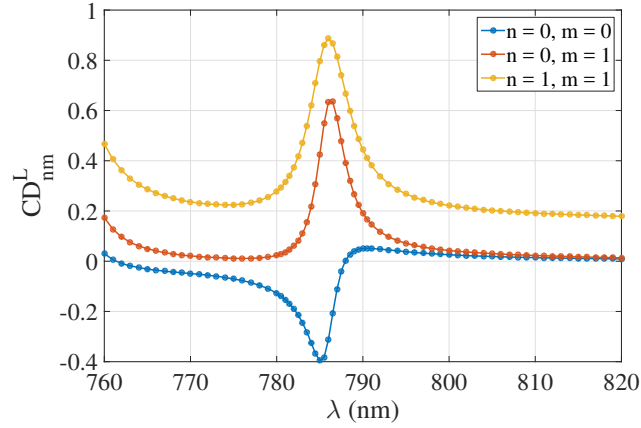

Fig. 7. Circular dichroism $CD_{nm}^L$ in transmittances (Eq. 13) versus vacuum wavelength for the square lattice made of left twisted nanoholes (Fig. 3). Results for three diffraction orders, namely, $n$=0, $m$=0 (blue line, standard CD), $n$=0, $m$=1 (red line) and $n$=1, $m$=1 (yellow line), are shown.

### 3.2 Circular dichroism in transmission

One may expect that the system under consideration would exhibit circular dichroism (CD). One can characterize this CD in transmission for all diffraction orders in natural way, i.e.,

$$CD_{nm}^L = \frac{T_{R \to R, nm}^L - T_{L \to L, nm}^L}{T_{R \to R, nm}^L + T_{L \to L, nm}^L}, \quad (13)$$

where $T_{R \to R, nm}^L$ and $T_{L \to L, nm}^L$ are given by Eq. (10). In Fig. 7, the wavelength dependence of the calculated $CD_{nm}^L$ for all considered diffraction orders for the square lattice of left twisted gammadions (Fig. 3) is displayed. Figure 7 shows clearly that our system exhibits CD and hence is a truly chiral one despite of a small difference in refractive index between the substrate (SiO$_2$) and superstrate (immersion oil) (see Fig. 3). This is interesting because in the case when substrate and superstrate have the same refractive indices, the system would possess mirror symmetry and hence exhibit no chiral property. Moreover, given the small thickness of the gold film (220 nm), the chirality and CD showed by our system is large.

Another useful definition of circular dichroism is related to the fact, that a linearly polarized incident light beam would become elliptically polarized after passing through the square lattice of chiral nanoholes. Therefore, circular polarization of transmitted light can also be characterized with ellipticity $\theta$ [26]

$$\theta = \arctan\left(\frac{|E'_{R,nm}| - |E'_{L,nm}|}{|E'_{R,nm}| + |E'_{L,nm}|}\right), \quad (14)$$

where $|E'_{R,nm}|$ and $|E'_{L,nm}|$ are, respectively, the magnitudes of the electric field vectors of the RCP and LCP light when a linear polarized wave irradiates the system. The usefulness of this classical definition is the fact that $\theta = 0$ corresponds to the absence of CD and chirality, while $\theta = 45°$ corresponds to the full absorption of one component and maximal chirality. In Fig. 8, the wavelength dependence of the calculated $\theta$ for all considered diffraction orders for the square lattice of left twisted gammadions (Fig. 3) for light polarized along *y*-axis is displayed. Since ellipticity is defined for linearly polarized incident light, ellipticity in the diffraction orders with fixed $n^2 + m^2$ is no longer identical, as was the case for circularly polarized incident light. Now only diffraction orders $n, m$ and $-n, -m$ have equal ellipticity.

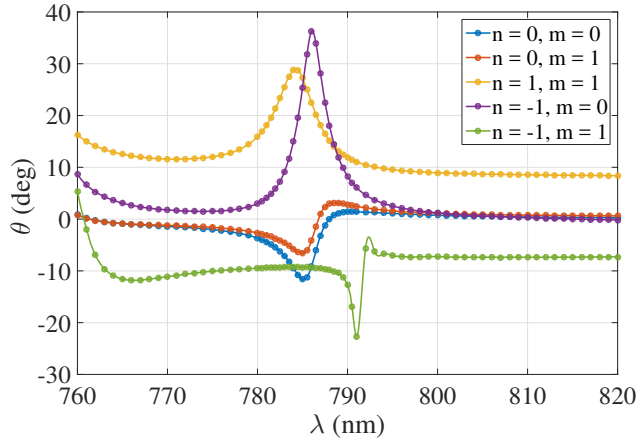

Fig. 8. Ellipticity $\theta$ in transmittances (Eq. 14) versus vacuum wavelength for the square lattice made of left twisted nanoholes (Fig. 3). Results for five diffraction orders, namely, *n*=0, *m*=0 (blue line, standard CD), *n*=0, *m*=1 (red line), *n*=1, *m*=1 (yellow line), *n*=-1, *m*=0 (purple line) and *n*=-1, *m*=1 (greed line), are shown. The system is irradiated by light that is linearly polarized along *y*-axis.

One can see from Fig. 8 that signs of $\theta$ are opposite for zero and nonzero diffraction orders. This means that incident linearly polarized wave would be converted into elliptically polarized waves with opposite helicities in zero and nonzero diffraction orders. In the present case, non-diffracted wave has left-elliptical polarization while diffracted waves are right-elliptically polarized. The ellipticity has the maximum values of 11.7° for zero diffraction order (for wavelength 785.2 nm), 36.5° for $n = -1, m = 0$ order (for wavelength 786 nm), 29° for $n = 1, m = 1$ (for wavelength 784.2 nm), 6° for $n = 0, m = 1$ (for wavelength 785.1 nm) and 22.5° for $n = -1, m = 1$ (for wavelength 791 nm). These large angles confirm again that our system has rather strong chirality. Moreover, further optimizations of the system could result in even stronger ellipticity.

*3.3 Spatial distribution of transmitted power and polarization state*

To understand an overall power distribution over different diffraction channels, we display polarization state for each diffraction maximum for two wavelengths $\lambda = 765$ nm and $\lambda = 786$ nm in Figs. 9(a) and (b) for parameter $s = 86$ nm (Fig. 4). It is clear from Fig. 9(a) that for wavelength $\lambda = 765$ nm, power diffracted to a side channel of the first-order is roughly equal to 1.2 %, while power going into the principal lobe is 2 times larger. Also, the total power transmitted to all side lobes is more than 3 times greater than power going into the main lobe. Another interesting feature is that polarization states of all orders with $n^2 + m^2 = 2$ are purely linear despite of LCP incident wave [Fig. 9(a)]. In contrast, Fig. 9(b) shows that for $\lambda = 786$ nm, nearly no energy is transmitted to $n^2 + m^2 = 1$ diffraction orders, while roughly equal amount of energy goes to the zero diffraction order and nonzero diffraction orders of $n^2 + m^2 = 2$. Remarkably, polarization of the waves in diffraction orders of $n^2 + m^2 = 2$ is almost purely RCP, while that of the zero diffraction order is the same as that of incident wave, as it should [Fig. 9(b)]. The energy efficiency of the polarization conversions reported in Sec. 3.1. is not high (i.e., about 1% of the incoming energy). Nevertheless, we expect that optimizations by making use of optical Tamm states can enhance this efficiency substantially [27-28].

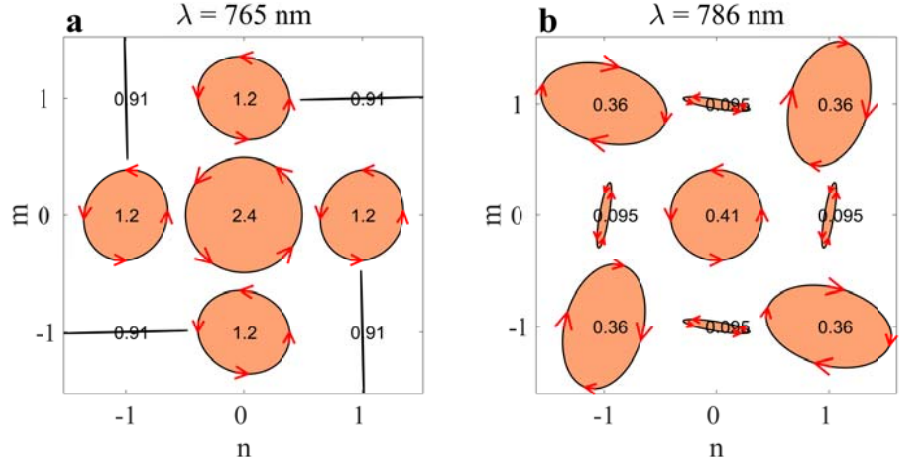

Fig. 9. States of polarization and transmittances (Eq. 10) (in percentages) of all considered diffraction orders for LCP plane waves of wavelengths $\lambda = 765$ nm (a) and $\lambda = 786$ nm (b), incident on the square lattice of left twisted nanoholes (Fig. 3). Arrows indicate rotation directions of the electric field vector (helicity). Clockwise rotation corresponds to RCP and counterclockwise rotation corresponds to LCP. In fact, this figure also represents spatial distribution of polarization and transmittance.

It would also be important to know relations between transmittances to different diffraction orders. To this end, one can define transmittance ratios

$$q^L_{P_1,2/0} = \frac{T^L_{P_1 \to R,11} + T^L_{P_1 \to L,11}}{T^L_{P_1 \to R,00} + T^L_{P_1 \to L,00}}, \quad q^L_{P_1,1/0} = \frac{T^L_{P_1 \to R,10} + T^L_{P_1 \to L,10}}{T^L_{P_1 \to R,00} + T^L_{P_1 \to L,00}}, \quad q^L_{P_1,2/1} = \frac{T^L_{P_1 \to R,11} + T^L_{P_1 \to L,11}}{T^L_{P_1 \to R,10} + T^L_{P_1 \to L,10}}, \quad (15)$$

where upper index L denotes twist of gammadions and $P_I (= \text{R or L})$ denotes polarization state of incident wave. The wavelength dependence of these ratios for the array of left twisted nanoholes for LCP and RCP incident waves are shown in Figs. 10 and 11, respectively.

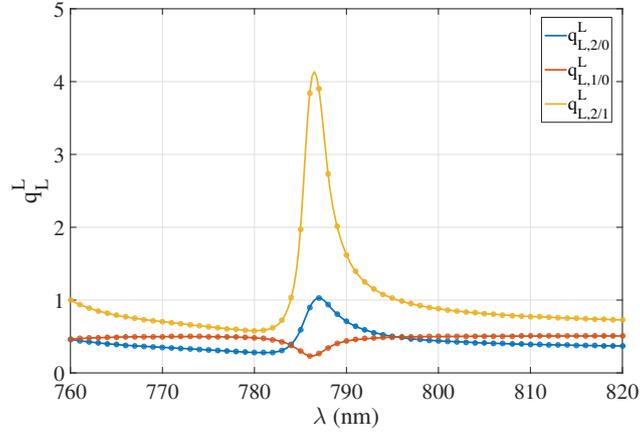

Fig. 10. Relative transmittances $q^L_{L,2/0}$, $q^L_{L,1/0}$, $q^L_{L,2/1}$ (Eq. 15) versus wavelength, of a LCP wave incident on the square lattice of left twisted nanoholes (Fig. 3).

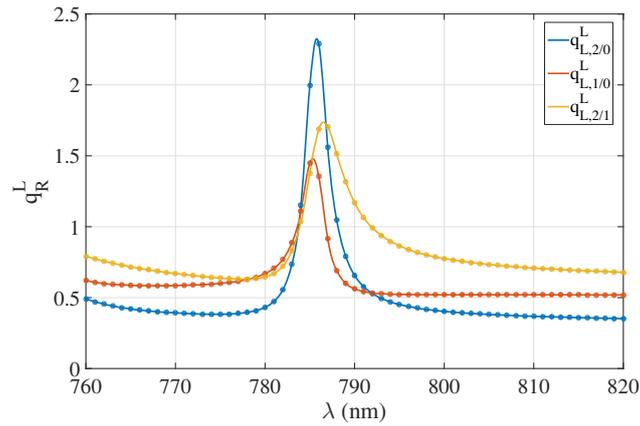

Fig. 11. Relative transmittances $q^L_{R,2/0}$, $q^L_{R,1/0}$, $q^L_{R,2/1}$ (Eq. 15) versus wavelength, of a RCP wave incident on the square lattice of left twisted nanoholes (Fig. 3).

Figure 10 shows that when helicity of incident wave is the same as that of gammadions, the average transmittance to zero diffraction order is about twice as large as the transmittances to orders $n = 1, m = 0$ and $n = 1, m = 1$ with exceptions at wavelengths around 785 nm. For wavelengths in the vicinity of 785 nm, transmittance to $n = 1, m = 1$ diffraction order is the same as transmittance to zero diffraction order. However, the peak of $q^L_{L,2/1}$ near this wavelength is not interesting because transmittances to both $n = 1, m = 1$ and $n = 1, m = 0$ diffraction lobes are very small (Fig. 4). Interestingly, situation is different when the twists of incident wave and gammadion are opposite (see Fig. 11). Although average transmittance into zero diffraction order is still approximately two times larger than transmittance to order

$n=1, m=1$ ($q_{R,2/0}^L \approx 0.5$), power redistributes significantly among different diffraction orders in a narrow wavelength range (782-792 nm) and, for wavelength 785, in particular, $q_{R,2/0}^L \approx 2.5$. The $q_{R,1/0}^L$ and $q_{R,2/1}^L$ show similar behaviors.

We have also carried out the calculations for light entering the system from the immersion oil side. Nevertheless, in this case, the system exhibits similar properties of polarization conversion, although the polarization conversion is weaker. In particular, the maximum value of the asymmetry coefficient [Eq. (12)] is 0.7 and is slightly smaller than 0.9 for the case of light incident from $SiO_2$ side. Also, the corresponding total transmittance to $n^2 + m^2 = 2$ orders is only 0.3 %, being significantly smaller than 1.4 % for light incident from $SiO_2$ side.

## 4. Conclusions

A formulation for analysis of polarization conversion in different diffraction orders in a square lattice of chiral nanoholes has been developed. It allows us to calculate polarization dependent transmittance to each diffraction channel. Generalization of this formulation to other types of planar lattices (e.g., hexagonal lattice) is straightforward. Application of this formulation to a planar array of gammadion-shaped nanoholes in a gold film reveals that CD ellipticity in non-zero diffraction orders can be strong (up to 36.5°) and also that polarization conversion in non-zero diffraction orders can be effective. Indeed, in certain wavelength ranges, LCP incident light waves can be almost completely converted into RCP waves scattered into diffraction orders $n^2 + m^2 = 2$ and $n^2 + m^2 = 0$ by the planar array of left twisted gammadions. This interesting effect has been attributed to the interplay of radiation pattern of single chiral nanohole cell and diffraction pattern of a square lattice. Furthermore, it is found that significant power redistribution among different diffraction orders can be engineered by adjusting wavelength and polarization of incident light as well as period of metastructure and refractive indexes of surrounding media. These interesting findings suggest a novel route for developing optical devices made of planar chiral metafilms with controllable polarization state, spatial direction and intensity of outgoing light.

**Acknowledgments**

V. V. Klimov, I. V. Zabkov and A. A. Pavlov acknowledges financial support from Advanced Research Foundation (Contract No. 7/004/2013-2018) and also from the Russian Foundation for Basic Research (Grants No. 14-02-00290 and No. 15-52-52006). G. Y. Guo, R.-C. Shiu and H.-C. Chan thank the Ministry of Science and Technology as well as National Center for Theoretical Sciences of Taiwan for financial support.